\documentclass[final]{svjour2}
\usepackage{graphicx}
\usepackage{epstopdf}
\usepackage{rotating}
\usepackage{amssymb}
\usepackage{mathptmx}
\usepackage[numbers]{natbib}
\makeatletter
\journalname{Journal of Low Temperature Physics}

\bibpunct{}{}{,}{s}{}{,}

\begin{document}

\newcommand{\hdblarrow}{H\makebox[0.9ex][l]{$\downdownarrows$}-}
\title{Validation of Phonon Physics in the CDMS Detector Monte Carlo}

\author{K.A.~McCarthy$^{1*}$ \and S. W.~Leman$^{1}$ \and A.~Anderson$^{1}$ \and D.~Brandt$^2$ \and P.L.~Brink$^3$ \and B.~Cabrera$^3$ \and M.~Cherry$^3$ \and E.~Do~Couto~E~Silva$^2$ \and P.~Cushman$^4$ \and T.~Doughty$^5$ \and E.~Figueroa-Feliciano$^1$ \and P. Kim$^2$ \and N.~Mirabolfathi$^5$ \and L.~Novak$^3$ \and R.~Partridge$^2$ \and M.~Pyle$^3$ \and A. Reisetter$^{4, 6}$ \and R.~Resch$^2$ \and B.~Sadoulet$^5$ \and B.~Serfass$^5$ \and K.M.~Sundqvist$^5$ \and A.~Tomada$^3$ \and for the SuperCDMS Collaboration}

\institute{1: Massachusetts Institute of Technology,\\ Cambridge, MA, 02139, USA\\
(2) SLAC National Accelerator Laboratory,~Menlo Park, CA 94309, U.S.A. \\
(3) Stanford University, Stanford, CA, U.S.A. \\
(4) University of Minnesota, Minneapolis, MN 55455, USA \\
(5) The University of California at Berkeley, Berkeley, CA, U.S.A. \\
(6) Saint Olaf College, Northfield, MN 55057, USA \\
\email{*Corresponding author: kevmc@mit.edu}
}


\date{07/13/2011}

\maketitle


\begin{abstract}
The SuperCDMS collaboration is a dark matter search effort aimed at detecting the scattering of WIMP dark matter from nuclei in cryogenic germanium targets. The CDMS Detector Monte Carlo (CDMS-DMC) is a simulation tool aimed at achieving a deeper understanding of the performance of the SuperCDMS detectors and aiding the dark matter search analysis. We present results from validation of the phonon physics described in the CDMS-DMC and outline work towards utilizing it in future WIMP search analyses.
\end{abstract}

\section{Introduction}
The SuperCDMS phase of the Cryogenic Dark Matter Search (CDMS) experiment utilizes cylindrical germanium targets patterned with cryogenic Transition Edge Sensors (TESs) and ionization sensors in an effort to detect WIMP dark matter directly scattering from atomic nuclei \cite{Ahmed:2009zw, Ahmed:2010wy}. SuperCDMS detectors, known as iZIPs, consist of interleaved ionization sensors and TES phonon sensors with phonon-collecting aluminum fins patterned on both the top and bottom surfaces of a 1" thick germanium crystal \cite{2006NIMPA.559..414B, Oed1988351}. The phonon signal is read by multiple channels to provide information about event position. The CDMS Detector Monte Carlo (CDMS-DMC) simulates phonon and charge carrier propagation as well as the TES and ionization sensors to provide a new tool for investigating the behavior of the CDMS detectors. Tuning and validation of the simulated phonon physics is necessary to ensure that the CDMS-DMC accurately reproduces detector behavior. This validation is performed by comparing results from CDMS-DMC simulations with data taken at CDMS testing facilities and at the underground experimental site. 

\section{Distributions of Event Parameters}
\begin{figure}[ht]
\begin{minipage}[t]{0.5\linewidth}
\centering
\includegraphics[width=1\linewidth]{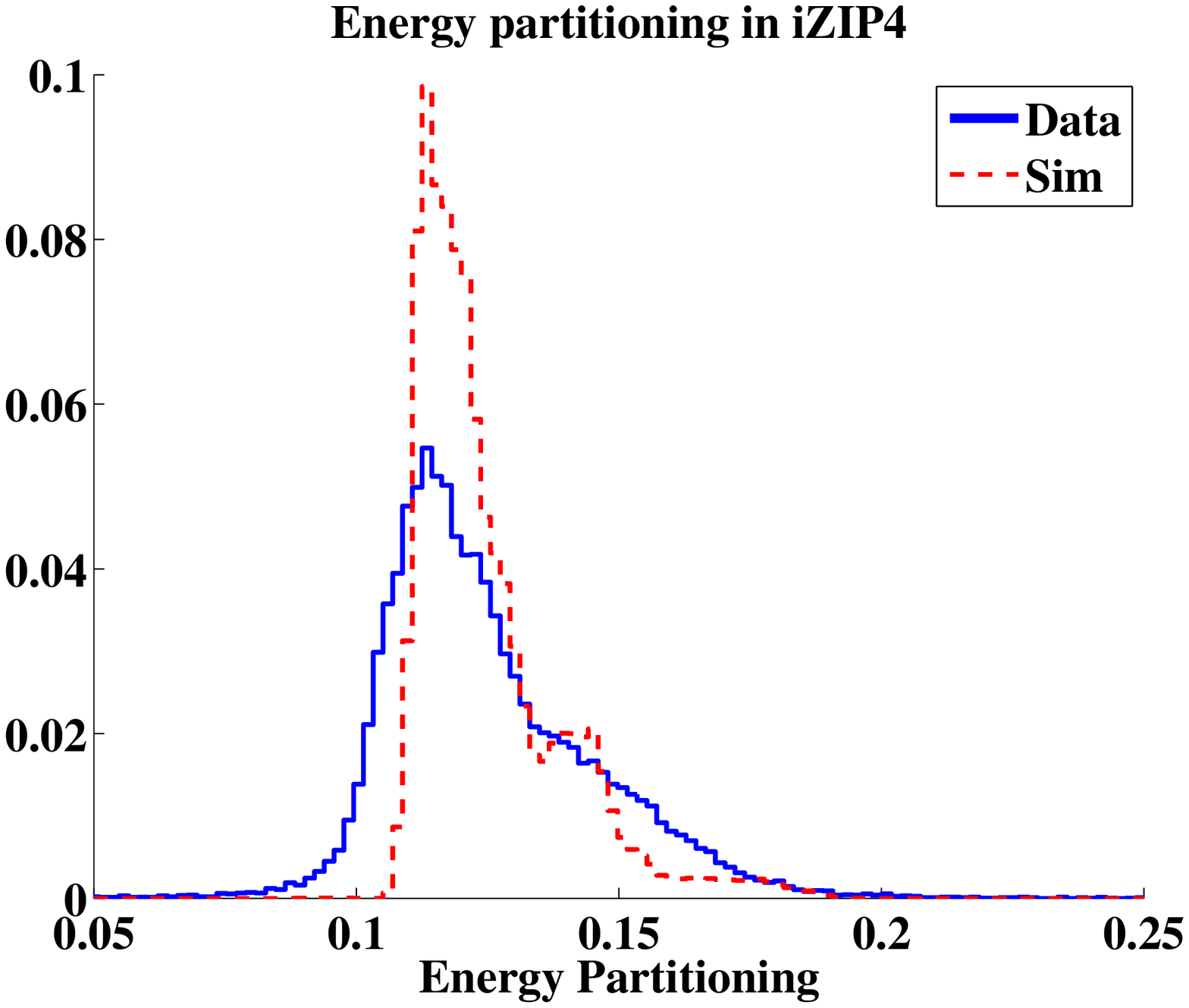}
\caption{(Color online) Distribution of energy partitioning (energy in one channel / total event energy) in real data events (solid blue) and CDMS-DMC simulations (dashed red) of an iZIP4 detector. The detector has 8 total channels, 4 on each detector surface, divided into 3 inner channels and 1 outer channels.  Each channel covers 1/4 of the detector surface. Each event contributes 8 entries to the histogram, one for each channel.  The y-axis represents the counts per bin normalized to sum to 1. The difference in widths indicates that the data exhibits slightly stronger energy partitioning than the simulation.}
\label{fig:Partitioning}
\end{minipage}
\hspace{0.5cm}
\begin{minipage}[t]{0.5\linewidth}
\centering
\includegraphics[width=1\linewidth]{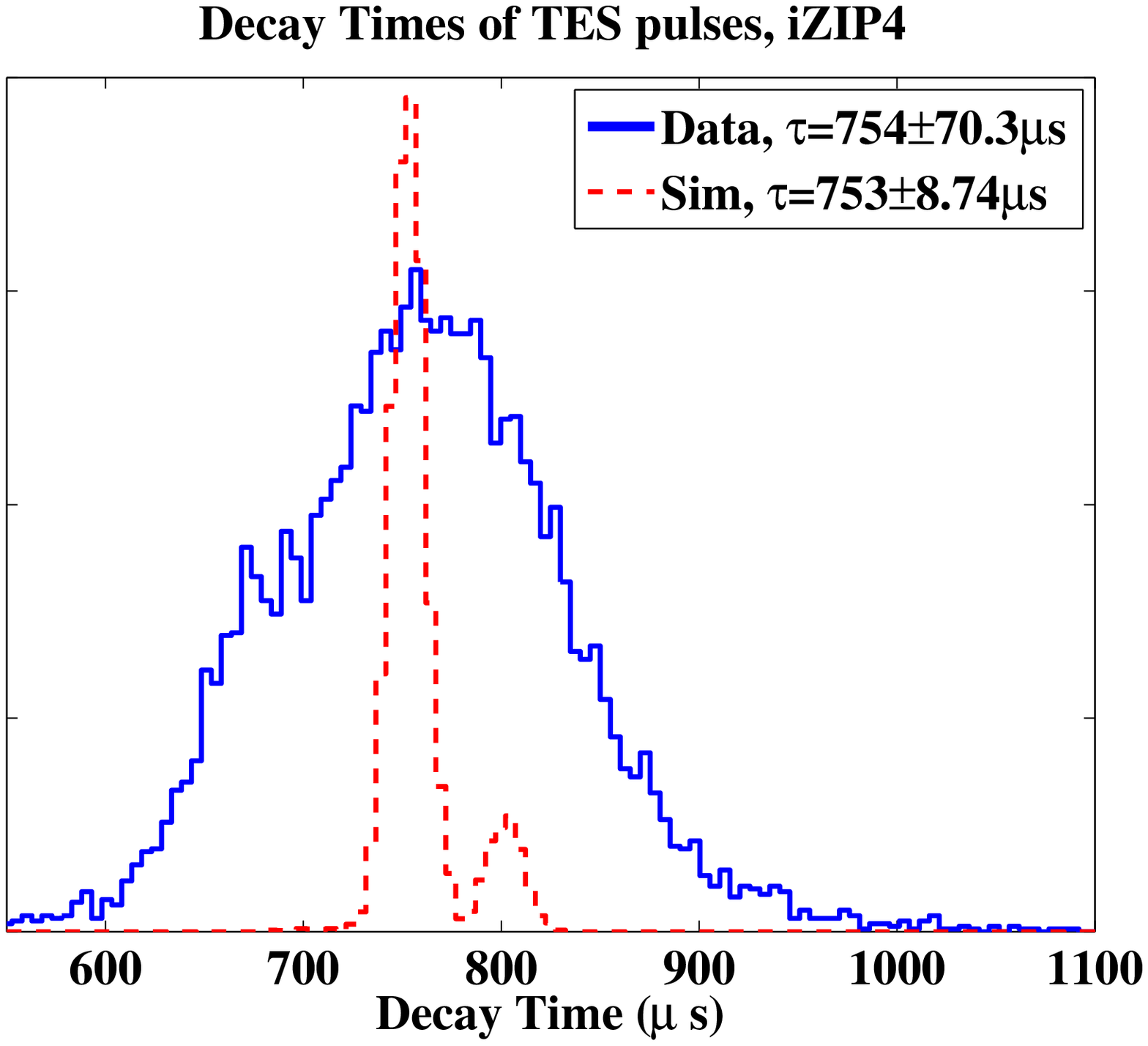}
\caption{(Color online) Decay times of TES pulses observed in data (solid blue) and in CDMS-DMC simulations (dashed red). The decay of each TES pulse is fit to a single falling exponential from 750 to 2750~$\mu s$ after the onset of the event. The decay time distributions are then fit to Gaussians to provide the means and widths indicated in the legend; these fits are not shown. The mean decay times of data and sim are in agreement; the differences in the widths of the distributions are most likely due to channel variations and noise. The secondary peak in the simulation histogram is due to a single TES channel. The y axes used for the two histograms are scaled relative to one another to allow for comparison.}
\label{fig:Decay}
\end{minipage}
\end{figure}

In recent years, the CDMS collaboration has produced a number of new detectors with various TES sensor layouts on the surface. This variety of detector topologies provides an opportunity to validate the phonon physics in the CDMS-DMC across multiple detectors~\cite{Leman2009, Leman2011_4}, which prevents `over-tuning' of physical parameters to effects stemming from particular detector layouts and allows for confirmation of the validation procedure. Much validation work has focused on the CDMS-DMC's ability to reproduce distributions of event parameters (energy partitioning into the different TES channels, pulse rise and fall times, etc.) produced when the detector is exposed to deeply-penetrating gammas from a calibration source. Figures  \ref{fig:Partitioning} and \ref{fig:Decay} show two example distributions from real data and simulations of events dispersed throughout the bulk of an iZIP detector with four TES channels on each surface, split into three inner and one outer channels. Figure \ref{fig:Partitioning} shows a histogram illustrating the partitioning of phonon energy into the 8 TES channels, while Figure \ref{fig:Decay} shows a histogram of the phonon pulse decay times, which are determined by how rapidly phonons are removed from the crystals by the phonon-collecting Al fins. Comparing these distributions and many others has allowed us to tune the probability of phonon absorption vs. reflection at the patterned surfaces, compare measurements of the TES critical temperatures, consider different values for the anharmonic decay constant (eventually converging to near the literature value), and validate the area of aluminum coverage on the detector surface used in the CDMS-DMC.

\section{Phonon Pulse Matching}
\begin{figure}
\begin{center}
\includegraphics[%
  width=0.75\linewidth,
  keepaspectratio]{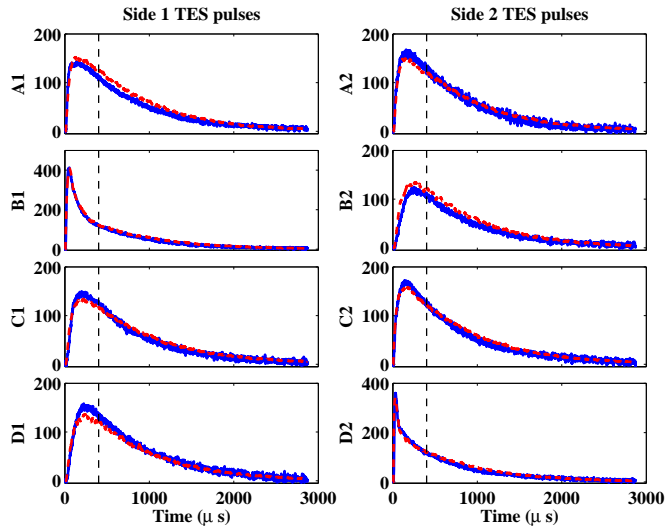}
\end{center}
\caption{(Color online) Set of phonon pulses from the 8 TES channels of the iZIP4 detector from a single data event (solid blue) shown with the best match found among the set of simulated events (dashed red). The data was taken at the UC Berkeley SuperCDMS testing facility. The left and right columns show pulses from the 4 channels on the top and bottom surfaces of the detector, respectively. Channel A is the outer TES channel, and channels B, C, and D are the 3 inner channels on each side. The y-axis represents power in the TES in arbitrary digitized units. The black vertical lines indicate the range (400~$\mu s$) of the zoomed figure shown below.}
\label{pulsematch}
\end{figure}

\begin{figure}
\begin{center}
\includegraphics[%
  width=0.75\linewidth,
  keepaspectratio]{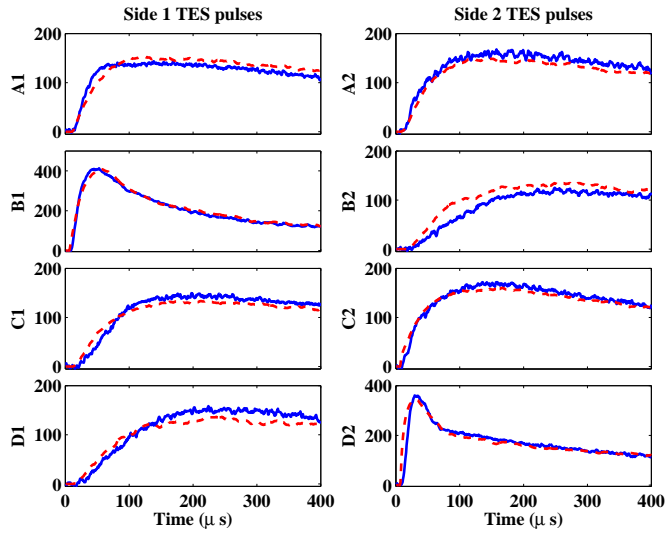}
\end{center}
\caption{(Color online) Zoomed version of figure above, showing a set of phonon pulses from a single data event (solid blue) shown with the best match found among the set of simulated events (dashed red). Pulses are zoomed in to the first 400~$\mu s$. }
\label{pulsematch_zoom}
\end{figure}

The next phase of validation of the CDMS-DMC involves directly matching the sets of TES pulses observed in real data events with sets of pulses from events simulated in the Detector Monte Carlo. Matching TES pulses observed in data to simulated pulses requires not just accurate phonon physics but also a well-tuned description of the TES parameters, which is described elsewhere in these proceedings \cite{Anderson2011, Anderson2011_2}. The procedure for matching the TES pulses begins with thousands of electron recoils simulated on a grid of points throughout the germanium crystal. A single data event is selected, and the set of phonon pulses is compared to the phonon pulses from each of the simulated events. Each simulated event is fit to the data event using a single amplitude scaling parameter. A weighted sum of the residuals between the data pulses and the simulated pulses is used as the goodness of fit metric; after a data event has been compared to all simulated events, the simulated event with the best (lowest) goodness of fit metric is selected as the best match for that particular data event. Figures \ref{pulsematch} and \ref{pulsematch_zoom} show the results of this matching routine for a single illustrative data event and shows that the selected simulated event matches the data pulses quite closely. Figure \ref{pulsematch} shows the entire sampling length of the TES pulses, $\approx$2850 $\mu s$, while Figure \ref{pulsematch_zoom} zooms into the first 400~$\mu s$ of the pulse, which contains most of the position-dependent variation in pulse shape.


\begin{figure}
\begin{center}
\includegraphics[%
  width=0.75\linewidth,
  keepaspectratio]{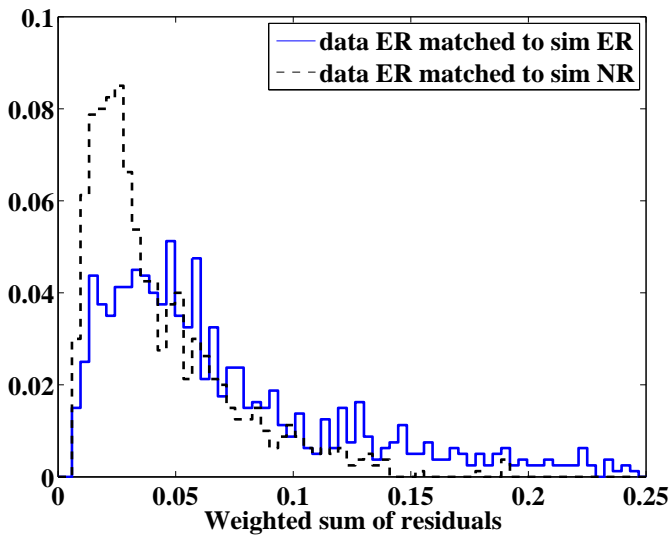}
\end{center}
\caption{(Color online) Distribution of the sum of weighted residuals (quality of pulse match) for bulk electron recoils from the data matched with simulated bulk electron recoils (solid blue) and simulated bulk nuclear recoils (dashed black). There is a complete lack of separation between the two distributions, and interestingly, the bulk NR simulations seem to provide better matches to real electron recoils than the simulated electron recoils do. The y-axis is the counts per histogram bin normalized to sum to 1.}
\label{ERNR}
\end{figure}

The ultimate goal of the TES pulse matching work is to demonstrate pulse-shape based discrimination of background electron recoils in the bulk and surface regions of the detector from the nuclear recoils that represent the WIMP-nucleon scattering signal. Figure \ref{ERNR} shows the weighted sum of residuals when bulk electron recoils in the data are compared with simulated bulk electron recoils and simulated bulk nuclear recoils. At present, these distributions are not well-separated, and in fact many of the electron recoils from the data find better matches among the simulated nuclear recoils than the simulated electron recoils. Thus, this pulse matching analysis does not yet demonstrate the ability to discriminate signal from background based solely on simulated pulse shapes.

\section{Future Work}
Validation of the CDMS-DMC is an ongoing process of continuous refinement. Additional work regarding phonon transport Monte Carlo simulation can be found in the proceedings~\cite{Brandt2011}. Future CDMS-DMC work aims to create a model-based WIMP search analysis method utilizing the pulse matching scheme presented above. A first step toward this goal involves quantifying the degree of observed variation in the phonon pulse shapes as the event position changes. These results can then be used to determine how reliably simulated pulses at different positions can be interpolated to achieve stronger agreement between simulation and data. The simulations of electron and nuclear recoils must also be independently compared to the data to determine whether the CDMS-DMC produces the excess `peakiness' of electron recoils demonstrated in other work \cite{Hertel2011}, as this feature will be critical to a WIMP search analysis based on pulse-shape discrimination using the CDMS-DMC.

\section{Conclusions}
The CDMS-DMC is able to reproduce many features of the phonon and TES behavior observed in real CDMS detectors. This agreement has been demonstrated using broad distributions of multiple event features such as energy partitioning and pulse shapes and also on event-by-event matching of sets of real TES pulses from CDMS detectors to pulses simulated using the CDMS-DMC. The goals of matching individual events from real SuperCDMS data with events simulated using the CDMS-DMC is to understand the position-dependent variation in the pulse shapes and to perform a WIMP search analysis using pulse-shape-based discrimination of background electron recoils from the nuclear recoil signal. 

\begin{acknowledgements}
This work is supported by the United States National Science Foundation under Grant No. PHY-0847342.
\end{acknowledgements}

\bibliography{./references}{}
\bibliographystyle{mystyle}

%
%
%

\end{document}